# Evolve with Your Research – Stepwise System Evolution from Document-driven to Fact-centric Research Data Management in Materials Science

Victor Dudarev[0000-0001-7243-9096], Alfred Ludwig[0000-0003-2802-6774]

Materials Discovery and Interfaces
Ruhr University Bochum
Bochum, Germany

**Abstract:** The digitalisation of research requires data management systems capable of supporting a broad spectrum of usage scenarios, ranging from document-oriented repositories to fully factographic environments. This paper introduces a methodological approach for the stepwise development of such systems, illustrated by the MatInf Research Data Management System (RDMS). The proposed framework combines a graph-based STAR paradigm—emphasising <u>S</u>tatefulness, <u>T</u>raceability, <u>A</u>im, and <u>R</u>esult—with the SET methodology, which enables systematic <u>S</u>tandardisation, <u>E</u>xtraction, and <u>T</u>esting of research data. Together, these principles provide a pathway towards FAIR-compliant data infrastructures, facilitating reproducibility, re-use, and integration of heterogeneous materials science data. By demonstrating the gradual consolidation of research outputs into unified datasets, this study highlights how adaptive RDMS design can support accelerated scientific discovery and enhance collaborative research in large-scale projects.

## Introduction

Information technologies have become indispensable across a growing number of fields from industry to science. Gains in efficiency of inference of knowledge from data is now often inseparable from the ability to store, analyse, and use large volumes of data in decision-making processes. This is particularly critical in scientific research, where the trajectory of future work must be guided by thorough analysis of past achievements and accumulated experience. Minimising duplication is essential, but beyond that, data analysis—augmented by machine learning or domain expertise—can provide valuable foresight in identifying promising research directions [1].

This is especially evident in materials science, which drives technological progress by delivering new materials that enable economic growth combined with sustainable development. Successful innovation in this field depends on drawing systematically on the extensive knowledge already amassed. To this end, a growing ecosystem of information systems—both domain-specific and general-purpose—has been developed to support data collection and organisation.

Research funding bodies increasingly recognise this need [2]. In fact, many now require research data management (RDM) as part of project design. For large-scale initiatives such as Collaborative Research Centres funded by the Deutsche Forschungsgemeinschaft [3], where typically more than a dozen research groups collaborate, an RDM system (RDMS) is a prerequisite [4]. Such RDMS provide a structured digital footprint of research activities, enabling reproducibility and data re-use, and thus extend the impact of research investments.

Recent advances in research data infrastructures such as NOMAD [5], Kadi4Mat [6], Chemotion [7], and Materials Data Facility [8] have considerably improved the accessibility and sharing of materials data in accordance with the FAIR principles [9]. Parallel developments in knowledge graphs—for example, the Materials Experiment Knowledge Graph [10] and MatKG [11]—highlight the growing importance of semantic representation and provenance tracking in materials science.

However, these initiatives primarily address infrastructural or data interoperability aspects, while methodological guidance on how to evolve a research data management system from document-oriented to fully factographic form remains limited. The present work fills this gap by introducing the methodological foundations for designing and evolving systems that function as shared information infrastructures for large-scale, collaborative materials science projects. These projects often assemble consortia that unite researchers from multiple disciplines and institutions to pursue common scientific goals. Because no single software solution can adequately support all forms of research data, the challenge lies in incrementally adapting and extending systems to consolidate diverse documents and datasets which are specific for the research project. We use our free open-source MatInf RDMS, which was developed within the project DFG CRC TR247, as a case study [12], demonstrating how it can be rapidly deployed for new research projects with baseline functionality and subsequently advanced towards tighter integration and consolidated research datasets.

**Document-Oriented Systems**

At the core of collaborative research in materials science lies the exchange of samples and scientific data among collaborating researchers (participants). In the early stages of any large-scale collaborative project, clear and detailed requirements are typically absent—whether concerning data formats, interaction processes among research groups, or the manner in which jointly obtained results should be represented. In practice, the launch of such projects almost always involves working with a variety of weakly structured data, where formats are determined either by the specifics of the used scientific and technical equipment or by standards that have emerged within individual subgroups of researchers. Every group can continue to use their own software systems, which are hopefully established to digitalise their research processes (if this is not the case, e.g. in a completely new research group, MatInf can be used to provide this basic functionality), e.g. Electronic Laboratory Notebook (ELN), Laboratory Information Management System (LIMS) or Research Data Management System (RDMS).

At first glance, it may seem that a document-oriented system, designed simply to record documents without analysing their internal content, would be ideal for handling such data in the initial phases. Such systems can be generic, deployed quickly, and sufficiently cover the immediate needs of even large research consortia in terms of information exchange. The minimum requirements typically include: (i) the ability to access the system from anywhere in the world without specialised software; (ii) information security (user authorisation, role-based access control, and differentiated levels of document access); (iii) the ability to structure projects (folders) for storing documents. These requirements are readily met by almost any of such systems.

What is fundamentally important at this stage, in the context of future system development, is the definition of *document types*. This classification enables documents to be categorised according to type and allows type-specific operations to be introduced for working with them, ensuring their meaningful use in subsequent research activities.

## Graph-Based Systems

In the short term, a document-oriented approach delivers rapid results. However, such systems provide limited possibilities for analysing stored documents and deriving added value from the data, i.e., it is difficult to infer knowledge from such data. In order to achieve this, RDMS must support the representation of relationships between stored documents—and ideally, these relationships should carry semantic information describing the nature of the connection. In computer science terms, this corresponds to graph-based systems, or systems capable of constructing a directed multigraph on top of existing documents, representing their interconnections (or, more generally, connections between objects).

For example, consider the link between a document describing a physical sample synthesised in the laboratory and another document reporting the measurement of its composition. The latter should also be linked to the measurement protocol (the procedure used) and to the measurement device (a document describing it). Essentially, a graph of related documents is obtained which can serve as a source of contextual information about the stored documents. While this context may or may not be obvious to the researchers directly involved in producing the documents, it is far less clear—or even incomprehensible—to external users. Without information on the sample, the procedure, or the measurement device, even an expert may struggle to assess the reliability of the data and the accuracy of the measurements. Thus, measurement protocols and device settings (parameters or metadata) are an integral part of the measurements themselves and must be reflected in the relationships between objects/documents.

For the sake of generalisation and conceptual clarity, we will henceforth use the term *object* rather than *document*. An object is defined as an entity in the database containing basic metadata (such as type, name, creation and modification dates, and author) and optionally a reference to an underlying document. In this way, every document corresponds to an object in the system, but not every object necessarily has a document associated with it—that is, objects without documents are also permitted.

## STAR Paradigm for Research Formalisation

To comply with the FAIR principles [9], the research process itself must be formalised at a high level and thoroughly documented. In the context of materials science [13], several aspects are of particular importance, which we summarise under the acronym **STAR**:

- **Statefulness** – the system must record every change in the state of a physical research object / sample (e.g., caused by processing or by specific measurements). This enables the association of measurement results with the correct state of the sample at different stages of its life cycle.

- **Traceability** – enables knowing the location of research samples (or the responsible personnel). This is particularly relevant in collaborative research projects (e.g., CRC), where samples may be repeatedly transferred between laboratories, making it necessary to track their movements or, at minimum, record the list of individuals who have worked with a sample. Additionally, traceability can include a second aspect: how closely the current state of the research aligns with the intended objectives, e.g., if the necessary samples were synthesised and characterised accordingly (Fig. 8).

- **Aim** – each experiment (or series of experiments) must have a clearly documented scientific objective (or aim), which could be formulated as a research question or hypothesis and has to be answered upon completion. To achieve this aim, a structured plan must be documented, comprising a sequence of tasks required to arrive at a substantiated conclusion. A dedicated object type, e.g., named *Idea or Experiment Plan*, may be used for this purpose. Such objects define the research aim and the plan in terms of object types that must be created as the research progresses. In this sense, an *Idea* functions as a root object, a container encompassing all objects/documents generated within its context. In graph-theoretical terms, the node representing the idea is connected to all subordinate nodes.

- **Result** – each research aim or plan must conclude with a report that summarises the outcomes of the experiments. It is very important that this can include both positive ("expected, wanted") and negative results. Such a report may take the form of a text document, a presentation, or a similar deliverable.

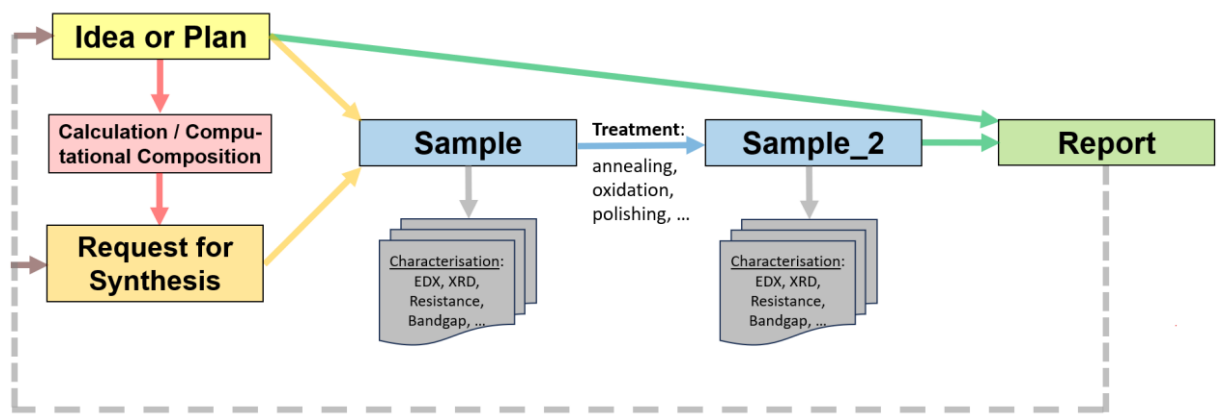

**Figure 1. STAR (Statefulness–Traceability–Aim–Result) paradigm as the cornerstone of research workflow.** STAR paradigm in research workflow: from an "Idea or Plan" to the result in a traceable way through stateful sample representation, starting from the design of an experiment plan (Idea, plan) and preparation for sample synthesis (optionally including theoretical justification), through state tracking of samples to ensure correct association of measurements with the relevant sample state, and concluding with a report of results. These results may provide grounds for continuing the research by formulating new aims, thereby closing the traceable research loop.

The STAR principle may be illustrated using a graph (Fig. 1) that depicts the research workflow. At the beginning of the research, the aim to be achieved or the hypothesis to be supported by evidence or to be falsified is formulated together with the measurement techniques that will support decision-making. All of this information is captured in the "Idea or Plan" object, which serves as the essential root object of the research. To proceed to the synthesis of specific samples, preliminary computations may be required (see "Calculation / Computational Composition"), resulting in a formal "Request for Synthesis" object. This specifies the desired compositions to be synthesised and sends notifications to the groups responsible for synthesis. As a result, a sample (or several samples) associated with the "Idea or Plan" object is produced and analysed according to the research plan. Any treatment or modification of the sample is recorded as a dedicated sample state object, linked to the characterisation documents referring to that state. The outcome of the research is summarised in a "Report" object, which contains the overall findings. Importantly, the conclusions presented in a report may themselves serve as the basis for new hypotheses and research plans; in this way, reports may be linked to new research cycles, closing the loop.

In a graph-based system built upon the STAR principle, isolated objects (nodes) should not exist. If they do, the rationale for their exclusion becomes questionable in terms of connectedness and proper documentation. The number of edges must not be smaller than the number of objects. Strictly speaking, a necessary (but not sufficient) condition is that the number of edges plus one should be at least equal to the number of objects.

Graphs constructed on top of documents undoubtedly help to organise and systematise information by assembling a coherent picture from individual entities. If semantics can be defined for different edge types, such a directed multigraph may be considered a **knowledge graph**, provided that inference rules are established over the sets of vertices and edges.

### Graph Edges Management

The management of relationships between objects in a graph-based system deserves particular attention, especially when the objective is to enable analyses grounded in the examination of directed links between objects. In other words, there must be strict rules governing the types of relationships allowed between objects of different types, taking into account their direction.

The specification of admissible edge types across the set of object types can be expressed through a relation **R** defined on the set of object types (the vertices of a graph **G = (V, E)**). Let $T_V$ denote the set of object types (vertex types) and $T_E$ the set of edge types. The relation **R** is then defined as a subset of the Cartesian product $R \subseteq T_E \times T_V \times T_V$. Depending on policy, **R** may represent either a whitelist (what is not explicitly allowed is forbidden) or a blacklist (what is not explicitly forbidden is allowed).

Thus, a triple $(t_E, t_{V1}, t_{V2}) \in R$ refers to a rule concerning edges of type $t_E$: it permits (or prohibits) edges of this type originating at an object of type $t_{V1}$ and terminating at an object of type $t_{V2}$. By enforcing such rules through triggers—mechanisms available in both relational and graph database management systems—it is possible to ensure the integrity of relationships in accordance with the specification defined by **R**.

### Factographic Systems

RDMS can be elevated to a new level through a stepwise transition—driven by the progressive refinement of requirements for document formats—towards **factographic RDMS** based on the strict formalisation of research data. Ideally, standardisation, together with metrics derived from research documents, should make it possible to compare objects of the same type with one another and to use these as features influencing the target properties under study.

For example, in electrocatalysis, when assessing catalytic activity, a crucial metric is the electrochemical surface area (ECSA) where the reaction takes place. This parameter enables the normalisation of current values, producing current densities that are suitable for comparison.

In this sense, starting from a graph of research documents, it is logical to identify the most important object types and initiate a stepwise transition to a factographic RDMS by formalising their data schemes in the form of relational tables or typed dictionaries (key–value pairs) (Fig. 2).

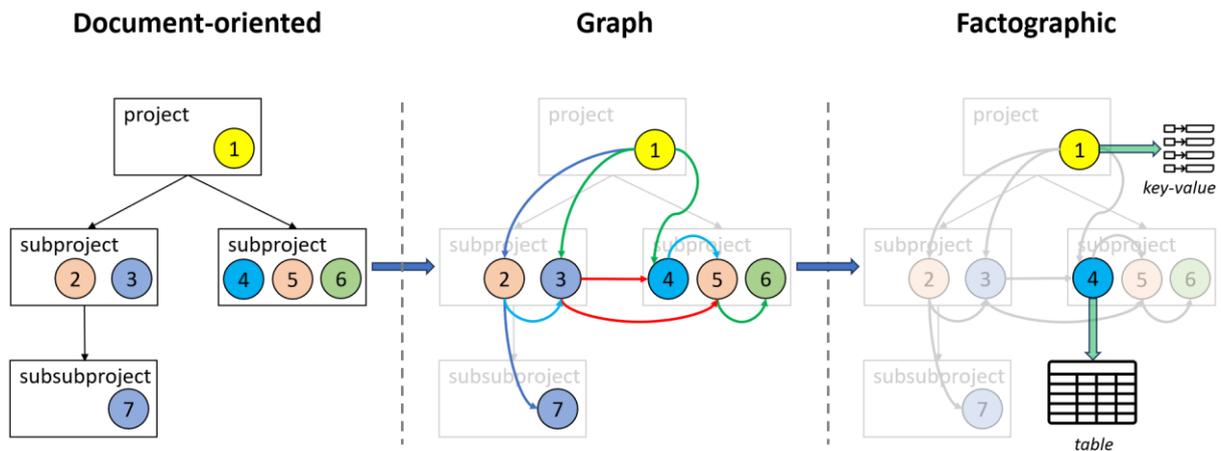

**Figure 2. Methodology for stepwise enhancement of RDMS based on the formalisation of object types.** Document-oriented RDMS provide only the bare minimum: documents structured in folders do not contain sufficient information about the relations between them. Graphs built on top of objects can be considered a significant improvement, especially if edges carry semantic information that allows the system to evolve towards Knowledge Graphs. The highest level of formalisation can be achieved in factographic systems, which can be gradually developed on top of graph systems by standardising object types and the data extracted from them, either in tabular or dictionary (key–value) form. Only factographic systems can serve as a source of integrated datasets consolidated from documents. Therefore, the gradual transition towards factographic RDMS, through continuous standardisation of object types, should be regarded as the ultimate aim.

The transition to factographic data for each document type (or, more broadly, object type) consists of three essential steps:

1) **Document format standardisation (Standardise)**. At this stage, a formal specification of the document is developed, describing its data structure and format. Special attention must be given to permissible value ranges and other constraints that valid data must satisfy. Target data—those intended for extraction and subsequent loading into the RDMS in factographic form—must also be specified, including the mapping of these data to specific object properties or relational table attributes. It is often necessary to support multiple formats of documents within a single object type, for instance when data are produced by measurement devices from different manufacturers.

2) **Validation of documents (Test / Validate)**. Based on the standardised format(s) developed in the first step, an algorithm is created to verify whether a document conforms to the required specification. The outcome of validation is a Boolean result (true/false), indicating whether the document complies with the format and whether its data can be reliably extracted. Only documents that successfully pass validation may be uploaded into the RDMS (either via GUI or API); otherwise, the system should reject the document. A generic object type, such as *Raw Document*, may be retained to allow the storage of unvalidated data, supporting iterative development of validation procedures and later reclassification into more specific types.

3) **Data extraction from documents (Extract)**. For validated documents, a procedure is implemented to extract factographic data and load them into the RDMS. The key limiting factor here is the expressiveness of the target RDMS, which must support for each object: (i) sets of typed key–value pairs, and (ii) tabular data of arbitrary structure. Thus, the extraction task reduces to the implementation of software modules capable of generating the necessary datasets for import into the RDMS.

This **SET methodology (Standardise – Extract – Test)** should ideally be applied to all object (document) types in the system, thereby transitioning it to a more standardised, rigorous mode of data handling. The reliability and automation of validation and data extraction procedures are the foundation for constructing datasets suitable for further analysis—one of the key characteristics of a mature RDMS.

Visualisation of standardised documents, like validation and extraction procedures, is carried out by external software components in line with microservice architecture principles. Flexibility in system expansion is achieved by configuring object types without modifying the RDMS core and by dynamically interacting with external services (Fig. 3).

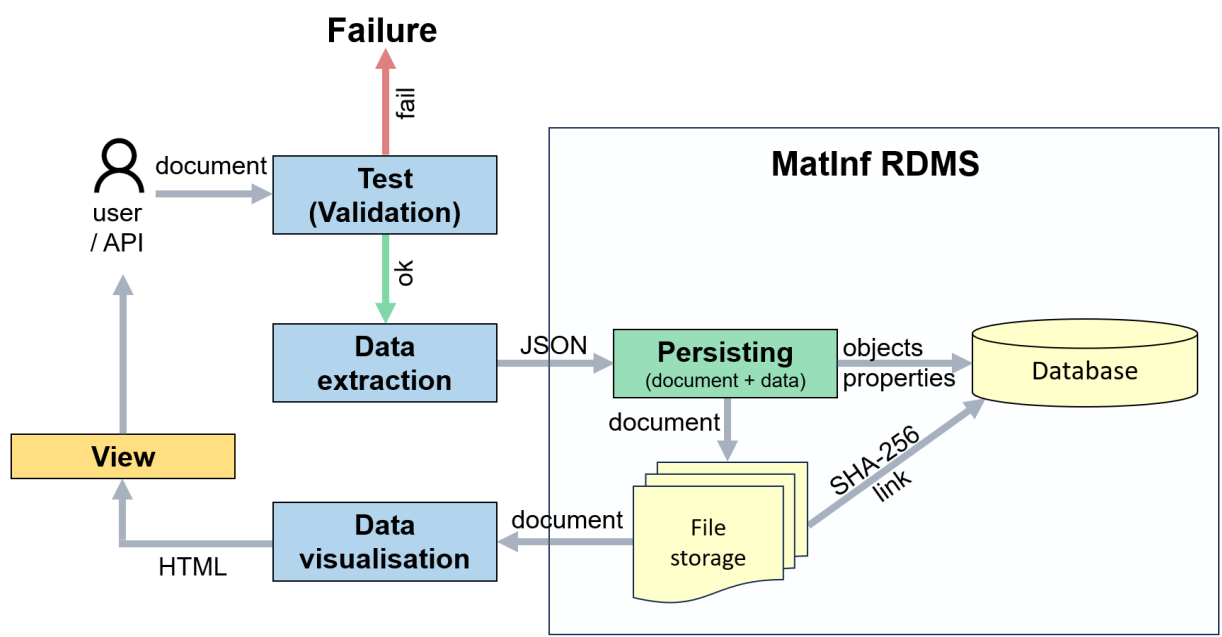

**Figure 3. SET-Methodology (Standardise–Extract–Test) for transitioning to factographic data using external web services for validation, data extraction, and visualisation.** Standardisation of an object type, together with the formalisation of dedicated document formats, provides the foundation for the deep integration of documents into MatInf RDMS. Deep integration refers to the automatic extraction of data (in tabular or dictionary form) from documents that have successfully passed validation for specification conformance. The testing, data extraction, and visualisation steps are handled by web services external to the RDMS. This architecture makes the solution modular and highly configurable: enabling deep integration with a new object type or file format requires only the deployment of supporting services and their configuration within the RDMS type settings.

It is worth noting that data extraction often requires contextual knowledge. For example, consider a document containing compositional measurement results (e.g. from EDX) of a thin-film sample (thickness ~100 nm) deposited on a substrate. Since the EDX method penetrates up to ~2 μm into the material, the measured composition will include contributions from both the film and the substrate. Thus, to obtain the quantitative composition of the thin film, the substrate contribution must be subtracted. This correction is possible because the EDX document is linked to its parent object (the thin-film sample), which specifies both the film composition and the substrate material. This illustrates how contextual information from higher levels plays a crucial role in correctly processing imported data.

## On the Data Storage Platform

Factographic systems are typically built on established relational database management systems (RDBMSs), which provide state-of-the-art functionality for working with tabular data, including referential integrity and transactional support. These mechanisms ensure a level of data consistency unattainable by most NoSQL solutions—an aspect of particular importance when the data are subsequently used for automated analysis.

At the same time, under pressure from document-oriented and graph-based approaches, traditional relational databases have evolved towards offering multimodal services. Most modern RDBMSs now include built-in support for graph data as well as for object-like data types (such as JSON and XML), which are more characteristic of document-oriented systems.

In this context, the choice of a modern relational database provides a solid and reliable foundation for the long-term development of RDMS. Such a choice not only ensures robustness in handling factographic data but also enables flexibility by supporting hybrid approaches that combine document-oriented, graph-based, and relational paradigms within a single RDMS.

## Methods / Experimental

Our objective was to develop an open RDMS for materials science that can flexibly accommodate a wide range of use cases—from document-oriented practices, typical of the early stages of building data management infrastructures, to factographic systems designed exclusively for standardized data and workflows. The ability to define and evolve appropriate data structures is central to enabling such flexibility.

In this context, we examine the three work scenarios outlined above, focusing on the core data structures that underpin them within the open-source **MatInf** system.

Document-Oriented Systems

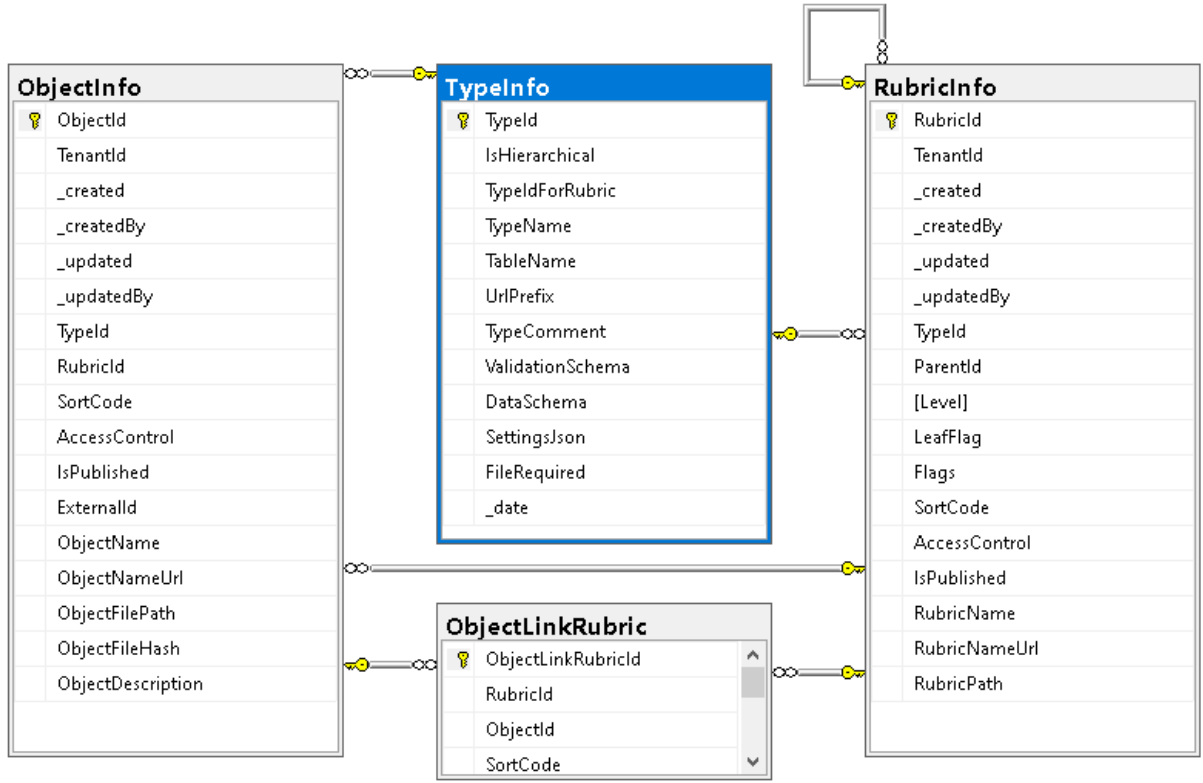

**Figure 4. Schema of the document-oriented part of the MatInf database.** Configurable object types are stored in the *TypeInfo* table, which forms the basis of a flexible type system and provides the foundation for the system's adaptability to any kind of data. The tree-structured project classifier (*RubricInfo*) enables the construction of a flexible project hierarchy, within which heterogeneous objects (*ObjectInfo*) can be contained. Additional links between objects and projects (*ObjectLinkRubric*) can be established, making objects accessible through multiple navigation paths.

Even at the document-oriented stage, it is crucial to ensure flexible typing of stored documents. In the MatInf system, this is achieved through an open, modifiable list of types stored in the *TypeInfo* table (Fig. 4). The concept of *type* in MatInf is used not only to characterise objects (documents), whose metadata are stored in the *ObjectInfo* table, but also to support hierarchical classifiers (projects and sub-projects), stored in the *RubricInfo* table.

Objects may also include cross-references to other projects, which is particularly important when an object of study must be associated with several projects simultaneously (via *ObjectLinkRubric* table).

As shown in Fig. 4, the schema for a generic document-oriented system is simple, while also providing sufficient flexibility for extension to materials science-specific types such as chemical systems and compounds, as described in [13].

Graph-Based Systems

When transitioning step by step to graph-based systems, the need for typed relationships between objects is addressed through the *ObjectLinkObject* table. In addition to metadata on links creation and modification, this table includes a reference to the object that characterises the link type via the *LinkTypeObjectId* attribute (Fig. 5).

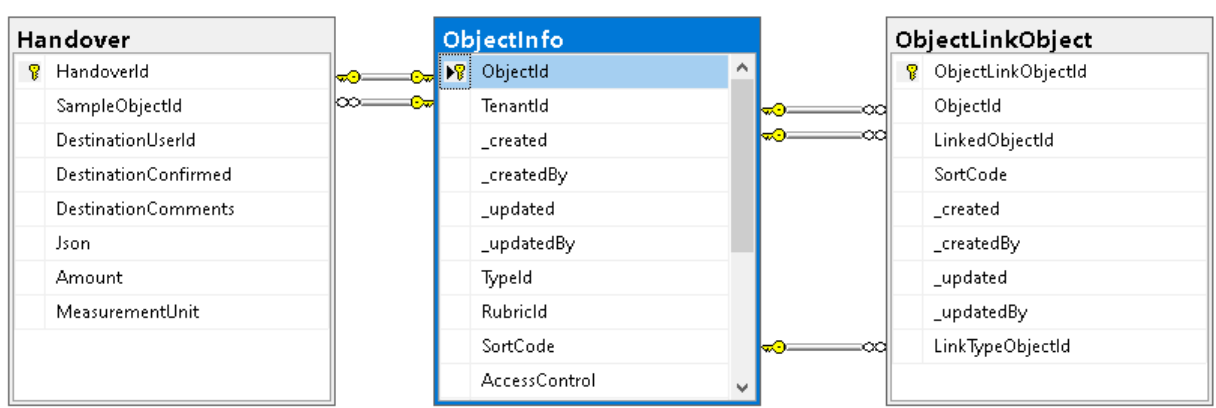

**Figure 5. Part of the database schema representing an oriented multigraph of objects and enabling sample traceability.** The *ObjectLinkObject* table is used to construct a directed multigraph, enabling the creation of many-to-many labeled relationships between objects. The *Handover* table contains information about the transfer of objects between research participants.

Based on such object relationships, graphs can be constructed to capture different states of research samples. An example of a graph tracking sample transformations in the laboratory—including reproduction, splitting, and annealing—is shown in Fig. 6.

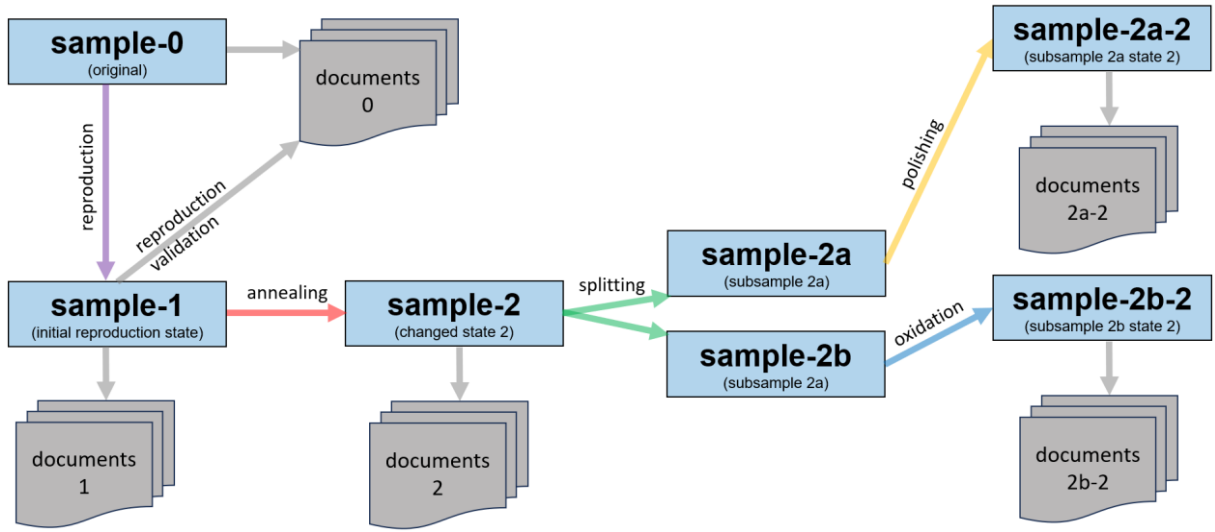

**Figure 6. Graph connecting sample states with documents containing measurement and analysis results at specific research stages.** To ensure traceability, it is crucial to record all changes in the state of research samples within the system and to associate measurement results with the corresponding state.

In large projects, the functionality to track the location of physical samples is of critical importance. This task is addressed through the maintenance of a handover register (*Handover* table in Fig. 5).

In MatInf, this functionality is implemented using dedicated objects of type *Handover*. The system allows the transfer of samples (any objects from the list of types permitted for handover), thereby enabling researchers to track their physical location and improve process efficiency through documented handovers, notifications, and reminders. The handover of samples proceeds in several steps (Fig. 7).

First, the sender ships the sample and records the transfer in the system. The recipient is notified by email and sees the pending transfer in their personal dashboard along with the sender's accompanying information. Upon receiving the sample, the recipient confirms the handover in the system, which automatically notifies the sender of completion. This mechanism ensures that the location of samples is always associated with the responsible researcher, enhancing transparency and accountability.

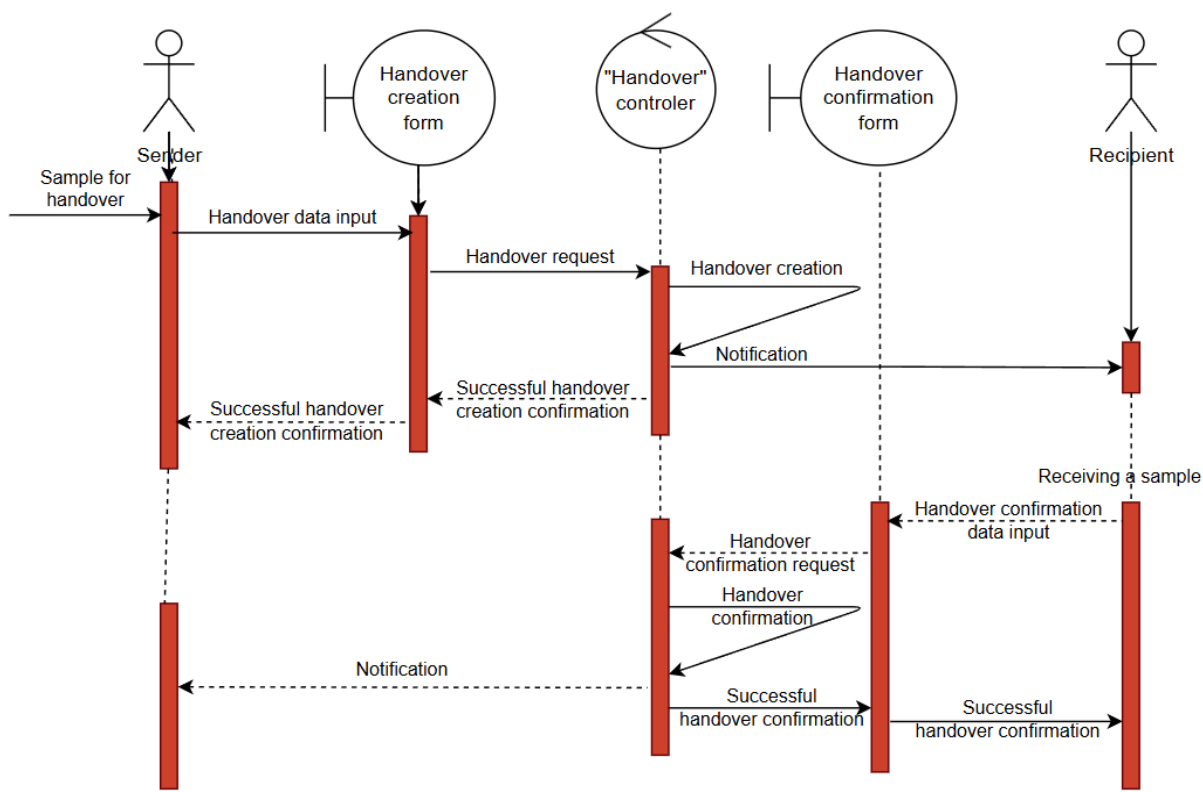

**Figure 7. Example of a sample handover with notifications.** When transferring a sample, the sender creates a handover object in the RDMS in the initial state, after which the recipient is notified of the upcoming transfer. Upon physically receiving the sample, the recipient confirms the receipt in the RDMS, thereby setting the *Handover* object into the completed state and notifying the sender of the successful delivery. By maintaining a list of such event objects, it is possible to efficiently track the sample's location and the current research stage.

In line with the STAR principles outlined above, it is essential to ensure that the research objective defined at the outset is continuously traceable, which is only possible through the formalisation of goal setting. In the MatInf system, when creating an object of type *Idea or Experiment Plan* (abbreviated as *Idea or Plan* in Fig. 1), the user may specify a list of object types—based on sample-related measurements—that must be associated with the studied objects as the research progresses. This constitutes a form of research plan specification. For example, in studying the catalytic activity of thin-film materials, the plan may include taking a photograph of the sample, performing compositional analysis (EDX), determining its crystal structure (XRD), and measuring electrical resistance as well as catalytic activity (high-throughput electrochemistry with SECCM) [14]. By recording measurements in the RDMS and associating them with investigated samples connected to the root object representing the research plan, the system automatically generates a tabular report for the given plan, illustrating the progress of ongoing research (Fig. 8).

| SampleId | ObjectName | N | System | SubstrateMaterial | Photo | EDX | XRD | Resistance | SECCM |
|---|---|---|---|---|---|---|---|---|---|
| 10269 | 10269 Ag-Au-Pd-Pt-Rh on 15nm Ta Library 1 | 5 | **Ag-Au-Pd-Pt-Rh** | Sapphire | 1 | 3 | 0 | 1 | 0 |
| 10275 | 10275 Ag-Au-Pd-Pt-Rh on 15nm Ta Library 2 | 5 | **Ag-Au-Pd-Pt-Rh** | Sapphire | 1 | 3 | 0 | 1 | 1 |
| 10304 | 10304 Au-Pd-Pt-Rh on 15nm Ta Library 1 | 4 | **Au-Pd-Pt-Rh** | Sapphire | 1 | 3 | 0 | 1 | 0 |
| 10311 | 10311 Au-Pd-Pt-Rh-Ru on 15nm Ta Library 1 | 5 | **Au-Pd-Pt-Rh-Ru** | Sapphire | 1 | 3 | 0 | 1 | 1 |

**Figure 8. Dynamic tabular report on the progress of the current research plan.** Red zeros indicate the absence of the corresponding objects and, consequently, the need to complete the respective stages of the study.

Factographic Systems

In transitioning to factographic systems—built on formalised data descriptions for each object type—the RDMS must be able to decompose data dictionaries or arbitrary structured tables into sets of object properties (scalar values) stored in dedicated *Property\** tables (Fig. 9). These tables share nearly identical structures, differing only in the type of values they store. The *PropertyFloat* table additionally includes a *ValueEpsilon* field, allowing the specification of measurement precision.

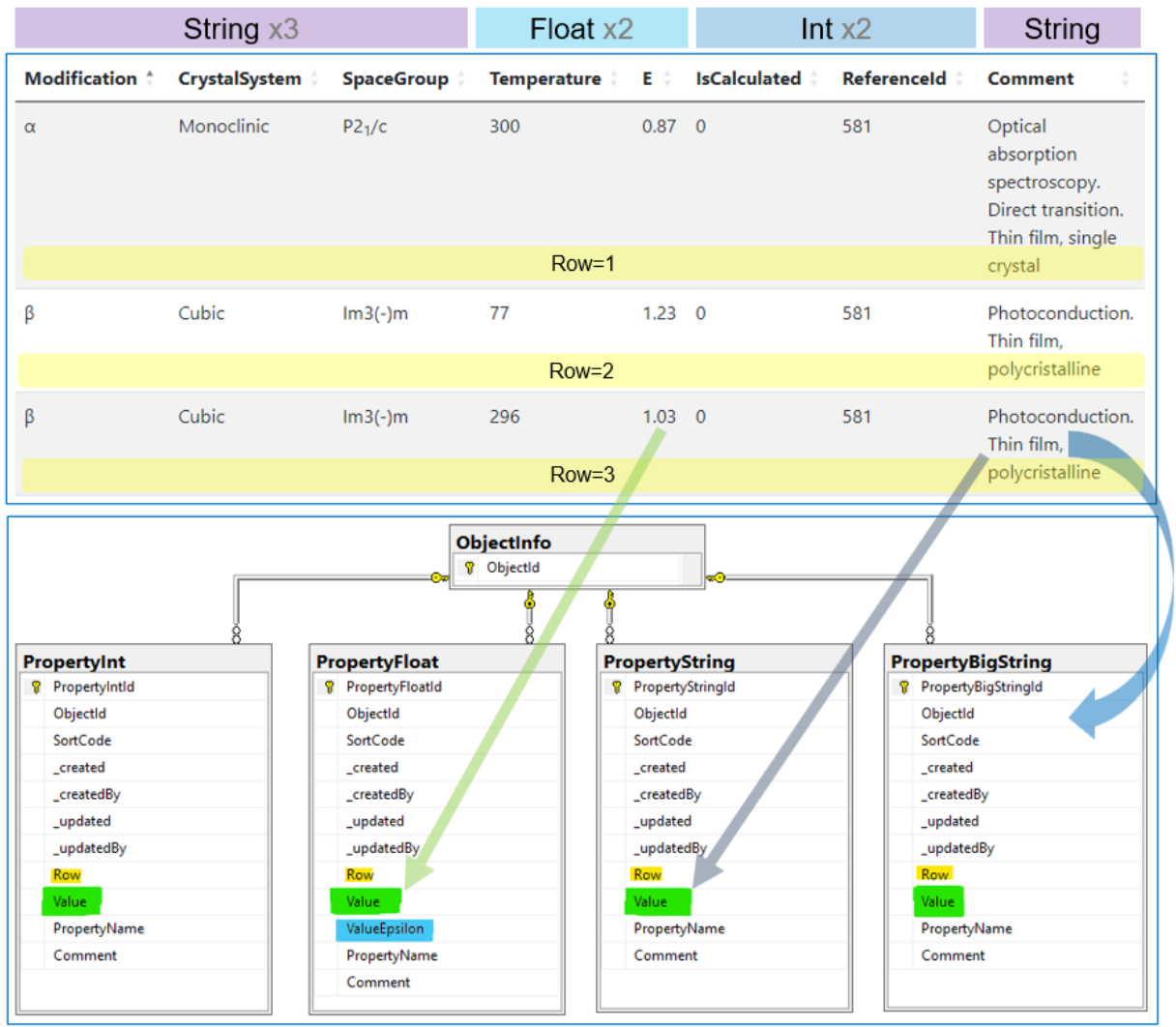

**Figure 9. Decomposition of arbitrary tables into property sets: example of band gap data.** The factographic part of the database equips objects of different types with configurable sets of properties and tabular data, where tabular data of any structure as well as typed key–value pairs are represented as records in the *Property\** tables.

Properties may be scalar (*Row is null*) or tabular (*Row > 0*), depending on the *Row* attribute. All extended property tables enforce a unique index on the triplet *(ObjectId, Row, PropertyName)*, which generalises to a unique quadruple *(TypeName, ObjectId, Row, PropertyName)* across property types. Consequently, an object cannot contain duplicate scalar properties of the same type. This constraint may be further strengthened, if required, by discarding *TypeName* and enforcing the rule directly at the database trigger level.

In this way, tabular data of arbitrary structure can be represented—for example, data on the variation of the bandgap width of $Ag_2S$ as a function of temperature and crystal structure (Fig. 9).

A distinctive feature of MatInf is the type-level configurable integration with external services (via a standardised API), which support validation, data extraction, and visualisation of documents. This allows services to be integrated incrementally, enabling a progressive transition to factographic data handling for the relevant object types, without modification of the MatInf core. Combined with the system's unique built-in capability for composition-range search of chemical entities [15]—particularly valuable for identifying compositionally similar materials—MatInf provides a versatile platform for the accumulation and processing of materials science data.

**Conclusions**

The digitalisation of research has made available a wide spectrum of software products, ranging from Electronic Lab Notebooks (ELNs) and Laboratory Information Management Systems (LIMS) to Research Data Management Systems (RDMS). Each of these solutions is designed with different emphases, and as such, they exhibit varying degrees of success in functioning as document-oriented repositories (potentially augmented with object-based graphs of interconnections) or as factographic research data management systems.

Document-oriented systems, when implemented without additional links between documents, generally fail to meet the broader **FAIR requirements** for RDMS. Even if partial contextual information is embedded within an individual document, connections to related documents or to the broader research context often remain implicit or entirely absent. Furthermore, the lack of explicit relationship management between documents makes it impossible to guarantee the logical integrity and connectivity of the stored information.

In our view, the minimally acceptable approach is a **graph-based system** implementing the **STAR paradigm**, as outlined above, through the use of typed document sets. A system designed in this manner has the potential to evolve in a structured way towards a **factographic system**, following the **SET methodology**. This approach enables the rigorous formalisation of research workflows—at the very least for specific stages, and in the ideal case for all aspects of research activities across object types. Such a system allows for the creation of datasets that integrate all relevant information (including synthesis parameters and measurement results) for arbitrary groups of samples.

The availability of such datasets would open the door to the discovery of new functional dependencies and significantly expand the possibilities of data analysis, thereby advancing research practices to a new level.

## Data availability

No special data is required to reproduce the results of this research–only the source code, which is publicly available as described in the code availability section below.

## Code availability

The source code of MatInf is available from https://gitlab.ruhr-uni-bochum.de/vic/infproject (MIT license).

Additional information is available from https://matinf.pro, public playground is available at https://inf.mdi.ruhr-uni-bochum.de.

## Acknowledgments

This research was financially supported by the Deutsche Forschungsgemeinschaft (DFG, German Research Foundation) Project-ID 388390466-TRR 247 (subproject INF) and CRC 1625, project number 506711657 (subproject INF).

## Author contributions

V.D. has been a major contributor to MatInf development and wrote the manuscript, A.L. supervised and coordinated the research, as well as supported writing of the manuscript.

## Competing interests

The authors declare no competing interests.